\documentclass[
 aip,jap,
 amsmath,amssymb,
 reprint,
]{revtex4-1}

\usepackage{graphicx}% Include figure files
\usepackage{dcolumn}% Align table columns on decimal point
\usepackage{bm}% bold math
\usepackage[utf8]{inputenc}
\usepackage[T1]{fontenc}
\usepackage{mathptmx}
\usepackage{etoolbox}

\usepackage{amsmath,amsfonts}
\usepackage[mathscr]{eucal}
\usepackage{array}
\usepackage{verbatim}
\usepackage{graphicx}
\usepackage[percent]{overpic}
\usepackage{siunitx}
\DeclareSIUnit{\dBm}{dBm}
\usepackage{multirow}
\usepackage{booktabs}
\makeatletter
\def\@email#1#2{%
 \endgroup
 \patchcmd{\titleblock@produce}
  {\frontmatter@RRAPformat}
  {\frontmatter@RRAPformat{\produce@RRAP{*#1\href{mailto:#2}{#2}}}\frontmatter@RRAPformat}
  {}{}
}%
\makeatother

\begin{document}

\preprint{AIP/123-QED}

\title[Coupled-Mode Equations with Arbitrary Mode Combinations]{Coupled-Mode Equations with Arbitrary Mode Combinations for Kinetic-Inductance Superconducting Traveling-Wave Parametric Devices: Theory and Experimental Validation}

\author{F.~P.~Mena}
\email{pmena@nrao.edu}
\affiliation{Central Development Laboratory, National Radio Astronomy Observatory, 1180 Boxwood Estate Rd, Charlottesville, VA 22903}

\author{R.~O.~Berriel}
\affiliation{Astronomy Department, Faculty of Physical and Mathematical Sciences, University of Chile, Camino El Observatorio 1515, Santiago, Chile}

\author{C.~Espinoza}
\affiliation{Central Development Laboratory, National Radio Astronomy Observatory, 1180 Boxwood Estate Rd, Charlottesville, VA 22903}

\author{R.~Finger}
\affiliation{Astronomy Department, Faculty of Physical and Mathematical Sciences, University of Chile, Camino El Observatorio 1515, Santiago, Chile}

\author{D.~J.~Thoen}
\affiliation{Space Research Organization Netherlands (SRON), Instrument Science Group - Litho, Niels Bohrweg 4, 2333 CA Leiden, The Netherlands}

\date{\today}

\begin{abstract}
The coupled-mode equations (CMEs) have proven to be very successful in describing different parametric processes in nonlinear optics. More recently, the same formulation has been used to model the behavior of microwave superconducting parametric amplifiers and frequency multipliers. However, when applied to the microwave regime, one has to take into account that not all the same assumptions remain valid and one expects that losses play a more dramatic role. Here, we revisit the CMEs applied to traveling-wave superconducting amplifiers not only to include losses, but also to provide a formulation that enables their systematic derivation for any combination of traveling waves. As examples, we discuss the impact of unwanted harmonics and intermodulation products on parametric amplification and harmonic generation. We verify that, if not taken into account properly, the performance of the devices can deviate considerably from the ideal case where the unwanted tones are not excited. Furthermore, by combining an independent experimental determination of $I'_*$ for a superconducting CPW artificial transmission line with simulations of its linear properties, we have obtained a parameter-free validation of this formulation. The nonlinear parameter was determined to be $I'_*\approx \SI{27}{\milli\ampere}$ which, surprisingly, scales with its theoretical depairing current and not with the actual much smaller critical current of the device. For the validation of the CMEs, we measured multiple-harmonic generation in the line and found an excellent agreement between theory and experiment up to the highest measured harmonic. This experiment also confirms independently the value of $I'_*\gg I_C$, which has direct implications for device design.
\end{abstract}

%\keywords{Superconductivity, amplitude equations, harmonic generation, CPW, artificial transmission line}

\maketitle

\section{Introduction}\label{sec:1_intro}
Their demonstrated capacity of achieving near quantum-limited noise has brought superconducting amplifiers considerable attention for their use in very sensitive applications like quantum computing or astronomy (for some recent examples see ~\cite{Splitthoff2024,Faramarzi2024,Sun2025,Aziz2025,Hao2026APL,Hao2026ARXIV}). In a broad sense, they can be classified into lumped-element and traveling-wave amplifiers. The former use individual superconducting elements such as JJs (Josephson junctions), SQUIDs (superconducting quantum interference devices) or SNAILs (superconducting nonlinear asymmetric inductive elements) to produce amplification. Due to their very nature, the main limitations of lumped-element amplifiers are low power handling and a limited operational bandwidth. These hurdles can, in principle, be overcome with traveling-wave versions using either a multitude of lumped elements connected in series or using the kinetic inductance of a long low-temperature superconductor transmission line~\cite{Fasolo2019,Aumentado2020,Esposito2021,Pagano2022}. The lumped versions are usually referred to as traveling-wave parametric amplifiers (TWPAs), while the latter, dubbed TKIPAs, have the additional benefit of possible operation at high frequencies, limited by the superconducting gap of the selected material (around~\SI{1}{\tera\hertz} in NbTiN)~\cite{Eom2012}, making them particularly attractive for applications in radio astronomy where they could be used at the RF or IF stages of heterodyne receivers. As is the case in most superconducting amplifiers, TKIPAs use the principle of parametric amplification, one of many possible parametric processes. Briefly, a parametric process is produced when different (cavity or traveling wave) modes present in a medium are coupled (through a non-linearity or excitation centers), resulting in an exchange of energy between them and, eventually, exciting ones whose initial amplitude was zero. If the correct conditions are created, large amplification of certain modes is possible.

Parametric processes can be studied using the coupled-mode equations (CMEs), also called amplitude equations, which can be obtained from either of two starting points. One is to start from the Hamiltonian of the used lumped-element device, leading to temporal-variation equations~\cite{Clerk2010,ROY2016}. The other is to consider propagating waves in a non-linear medium, through the wave equation, leading to spatial-variation equations~\cite{Yariv1973}. The description of TWPAs requires a link between these two frameworks~\cite{Fasolo2019}. On the contrary, since TKIPAs are constructed in the form of transmission lines, they can be studied entirely starting from propagating waves~\cite{Eom2012}. The term CMEs is usually applied only to the spatial-variation versions and this paper will follow that convention.

CMEs were first heuristically obtained for the study of multi-mode coupled microwave waveguides, and later introduced in the field of guided-wave optics to study coupled waveguides and nonlinear effects~\cite{Huang1994}. The CMEs have been amply studied in the field of non-linear optics~\cite{Boyd2008,Marhic2007,Agrawal2019} and the methods developed there can be almost directly applied to the study of TKIPAs~\cite{Eom2012}. However, an aspect not often discussed is the fact that not all of the assumptions used in nonlinear optics are applicable to the microwave range. Furthermore, there are two practical limitations when using the CMEs. (i)~They have to be deduced on a case-by-case basis depending on the number of traveling waves that are allowed to participate in the considered parametric process~\cite{Chaudhuri2015,Klimovich2024,Cunnane2024}. (ii)~The equations are normalized by the nonlinear parameter $I'_*$ preventing a parameter-free validation of the equations. Although not often discussed, it is used as a fitting parameter and usually taken to be approximately equal to $I_C$~\cite{Eom2012,Chaudhuri2015,Klimovich2024,Cunnane2024,Zmuidzinas2012}. In this work, we address these two limitations. First, we derive a general expression for the CMEs in superconducting transmission lines including losses and applicable to different parametric processes~(\S~\ref{sec:2_CMEs}). The formulation allows for the systematic derivation for any combination of traveling waves and is suitable for automated implementation. We then illustrate its use through examples, relevant to practical devices, including harmonic generation and four-wave mixing~(\S~\ref{sec:3_Examples}). We verify that, if high-order harmonics and intermodulation products are not taken into account properly, the desired response can deviate considerably from the ideal case where the unwanted tones are not excited. Second, we provide a parameter-free validation of the presented CMEs, in the case of harmonic generation, using a superconducting CPW artificial transmission line (\S\S~\ref{sec:4_Exp} and~\ref{sec:5_R&D}). In order to achieve this validation, we make an independent determination of $I'_*$ by combining simulations and measurements of the variation of the sample's RF transmission with injected DC current. The former are used to determine the properties of the line, namely the propagation constant and characteristic impedance, that are necessary to analyze the experimental results. We find, surprisingly, that the nonlinear parameter does not scale with the measured critical current but scales with the much larger theoretical depairing current, giving more precise design rules for device fabrication. Resistance-current measurements of the sample independently confirm this value and reveal the presence of dissipative regions whose onset is consistent with the harmonic generation results. The measured $I'_*$ allows us to pin the CMEs and make a direct comparison with a measurement of the harmonic generation in the line. We find a strikingly good agreement.

%explaining why very long lines and thin superconducting films are necessary to observe amplification in TKIPAs. 

%%%%%%%%%%%%%%%%%%%%%%%%%%%%%%%%%%%%%%%%%%%%%%%%%%%%%%%%%%%%%%%%%%%%%%%%

\section{Revisiting \& Extending the Coupled-Mode Equations}\label{sec:2_CMEs}
\begin{figure}[t]
    \centering
    \includegraphics[scale=0.4]{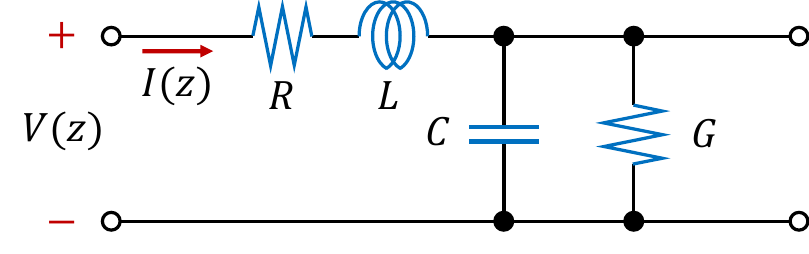}
    \caption{Lumped-element model of a transmission line considering losses. In the dispersion-engineered (DE) superconducting transmission lines considered in this work, the terms $R$ and $G$ come not only from the material properties but also from the presence of stopbands.}
    \label{fig:TL_model}
\end{figure}
As in~\cite{Carrasco2023} and~\cite{Longden2024}, we start from the more general model of a transmission line (Fig.~\ref{fig:TL_model}) and consider a superconductor medium. The superconductor is characterized by its inductance which is the sum of a geometrical and a kinetic parts~\cite{Anlage1989,Cho1997,Zmuidzinas2012},
\begin{equation}\label{eq:Ltot}
    L = L_G + L_{K0} \left[ 1 + \left( \frac{I}{I_*} \right)^2 \right]
    \equiv L_0 \left( 1 +\frac{I^2}{{I'_*}^2} \right),
\end{equation}
where $L_0=L_G + L_{K0}$, $I'_* = I_*/\sqrt{\alpha_K}$, and $\alpha_K = L_{K0}/L_0$ is the kinetic-inductance fraction. Notice that $I'_*$ is an important parameter determining the strength of the nonlinearity in a given device and has to be found experimentally. In order to obtain the CMEs, the next step is to obtain the wave equation using the telegrapher's equations.
\subsection{Linear Wave Equation ($I\approx 0$)}\label{sec:21-Linear}
First, we consider the case $I\approx 0$, where $L=L_0\neq L(I)$, leading us to a linear wave equation,
\begin{equation*}
    \mathcal{L}I\equiv \left( \partial_z^2 - C L_0 \partial_t^2 - (G L_0 + C R)\,\partial_t - G R \right) I = 0,
\end{equation*}
whose solutions are orthogonal linearly-independent attenuated plane waves, 
\begin{equation}\label{eq:plane_wave}
    \psi_m = e^{-\gamma_m z + j \omega_m t}.  
\end{equation}
The propagation constant and characteristic impedance at frequency $\omega_m$ are, respectively,
\begin{equation*}
    \gamma_m = \alpha_m + j\beta_m 
    = \sqrt{(R_m + j \omega_m L_{0m})(G_m + j \omega_m C_m)} 
\end{equation*}
and
\begin{equation*}
    Z_{Cm} = r_{Cm} + j x_{Cm} = \sqrt{\frac{R_m + j \omega_m L_{0m}}{G_m + j \omega_m C_m}}.
\end{equation*}

\subsection{Nonlinear Wave Equation ($I\neq 0$)}\label{sec:22-Noninear}
In the general case we obtain
\begin{equation*}
    \mathcal{L}I = \frac{1}{3}\left( G L_0 \partial_t + C L_0 \partial_t^2 \right) I^3\equiv \mathcal{N}I^3,
\end{equation*}
where we have made a variable change, $I \to  I/I_*'$. To solve this equation it is assumed that the current is a superposition of the plane waves given by Eq.~(\ref{eq:plane_wave}), but assuming a position-dependent amplitude,
\begin{equation}\label{eq:ansatz}
    I(z) = \frac{1}{2} \left[\sum_{n=1}^{N} A_n(z)\, e^{-\gamma_n z + j\omega_n t} + \text{c.c.} \right].
\end{equation}
In order to find the equation that governs the spatial evolution of any amplitude $A_m$, the next step is to project $\mathcal{L}I$ and $\mathcal{N}I^3$ over the temporal part of the plane wave $\psi_m$ using the ansatz~(\ref{eq:ansatz}). To facilitate the calculations, we define $\mathscr{A}_n(z) \equiv A_n(z)\, e^{-\gamma_n z}$ and $I_{\mathrm{c}} \equiv \sum \mathscr{A}_n(z)\, e^{j\omega_n t}$, which allow us to write the ansatz as $I = \frac{1}{2}(I_{\mathrm{c}} + I_{\mathrm{c}}^*)$ and its third power as $I^3 = \frac{1}{8}(I_{\mathrm{c}}^3 + 3I_{\mathrm{c}}^2 I_{\mathrm{c}}^* + \mathrm{c.c.})$. In this manner, the projection can be done over the operators applied to $I_{\mathrm{c}}$ and its powers, and, later, conjugate them. As an example, let us consider the first term of $\mathcal{L}I_{\mathrm{c}}$,
\[
\partial_z^2 I_{\mathrm{c}} = \partial_z^2 \textstyle\sum \mathscr{A}_n e^{j\omega_n t} = \textstyle\sum e^{j\omega_n t} \partial_z^2 \mathscr{A}_n.
\]
Its projection over the temporal part of $\psi_m$ is, then,
\[
\langle e^{j\omega_m t} | \partial_z^2 I_{\mathrm{c}} \rangle =
\textstyle\sum \, \partial_z^2 \mathscr{A}_n \langle e^{j\omega_m t} | e^{j\omega_n t} \rangle
= \partial_z^2 \mathscr{A}_m,
\]
where we have used the fact that $\langle e^{j\omega_m t} | e^{j\omega_n t} \rangle = \delta_{\omega_m,\omega_n}$. Proceeding in the same manner with all the other terms of the wave equation we obtain
\begin{align*}
&\frac{1}{2}\big(\partial_z^2 \mathscr{A}_m - \gamma_m^{2} \mathscr{A}_m\big) = -\frac{1}{24}\omega_m(\omega_m C_m L_{0,m} - jG_m L_{0,m}) \\
&\hphantom{\frac{1}{2}\big(\partial_z^2 \mathscr{A}_m - \gamma_m^{*2} \mathscr{A}_m\big)} \times\sum_{p,q,r} \big\{ \mathscr{A}_p \mathscr{A}_q \mathscr{A}_r \, \delta_{\omega_m, \omega_p+\omega_q+\omega_r} \\
&\hphantom{\frac{1}{2}\big(\partial_z^2 \mathscr{A}_m - \gamma_m^{*2} \mathscr{A}_m\big) \times\sum_{p,q,r} \big\{} +3\mathscr{A}_p \mathscr{A}_q \mathscr{A}_r^* \, \delta_{\omega_m, \omega_p+\omega_q-\omega_r} \\
&\hphantom{\frac{1}{2}\big(\partial_z^2 \mathscr{A}_m - \gamma_m^{*2} \mathscr{A}_m\big) \times\sum_{p,q,r} \big\{} +3\mathscr{A}_p^* \mathscr{A}_q^* \mathscr{A}_r \, \delta_{\omega_m, -\omega_p-\omega_q+\omega_r} \big\}.
\end{align*}
For practical use of this expression, we observe that
\[
-\omega_m(\omega_m C_m L_{0,m} - jG_m L_{0,m}) = j\frac{\gamma_m}{Z_{C_m}}\operatorname{Im}(\gamma_m Z_{C_m}) \equiv b_m,
\]
allowing to express it in terms of the characteristic impedance and propagation constant since they are directly accessible either experimentally or by simulations. Finally, we go back to the original variable for the amplitude, leading to
\begin{equation}\label{eq:CME}
    \begin{aligned}
        \mathfrak{A}_m &= \frac{1}{24} b_m e^{\gamma_m z} \\
        &\times \sum_{p,q,r=1}^{N} \Bigg\{
        A_p A_q A_r \, e^{-(\gamma_p+\gamma_q+\gamma_r)z}
        \, \delta_{\omega_m,\omega_p+\omega_q+\omega_r} \\
        &\qquad + 3 A_r^* A_p A_q \, e^{-(\gamma_p+\gamma_q+\gamma_r^*)z}
        \, \delta_{\omega_m,\omega_p+\omega_q-\omega_r} \\
        &\qquad + 3 A_p^* A_q^* A_r \, e^{-(\gamma_p^*+\gamma_q^*+\gamma_r)z}
        \, \delta_{\omega_m,-\omega_p-\omega_q+\omega_r}
        \Bigg\},
    \end{aligned}
\end{equation}
where $m = 1,\ldots,N$, and
\begin{equation*}
   \mathfrak{A}_m \equiv \frac{1}{2} A_m'' - \gamma_m A_m'.
\end{equation*}

Equation~(\ref{eq:CME}) represents a system of $N$ equations describing the spatial evolution of the $N$ modes assumed to be traveling in the superconducting transmission line. The Kronecker deltas serve to select the allowed combinations of the parametric process being studied, given the restrictions imposed by the modes allowed to travel. Therefore, this equation can easily be automatized to study any parametric process given a quadratic nonlinearity in the inductance, as in Eq.~(\ref{eq:Ltot}). An important aspect to consider is that the evolution of the amplitudes is only predicted as a fraction of $I'_*$, making its (experimental) determination crucial for practical applications.

Another important aspect of this equation is that it is more general than those presented in~\cite{Carrasco2023} and~\cite{Longden2024}. On the one hand, Carrasco \textit{et al.}~\cite{Carrasco2023} did not apply the projection over $\mathcal{N}I$ but over an approximation obtained by applying the method of multiple scales. On the other hand, Longden \textit{et al.}~\cite{Longden2024} did apply the projection over $\mathcal{N}I$ but applied only to a specific parametric process. Even more, they limited their study to cases with low losses where $\gamma \approx i\omega\sqrt{LC}\left[1-\frac{i}{2}\left(\frac{R}{\omega L}+\frac{G}{\omega C}\right)\right]$.

\subsection{Some Limiting Cases}
Before presenting examples on the use of Eq.~(\ref{eq:CME}), we discuss a few limiting cases:

\paragraph{Slowly-varying envelope approximation (SVEA):}
This is the usual assumption of a slowly varying wave resulting effectively in neglecting $A_m''$ in $\mathfrak{A}_m$.

\paragraph{Purely real impedance ($x_{Cm} = 0$):}
It results in $b_m = j \beta_m \gamma_m$, implying that, in this case, the CMEs do not depend on the impedance of the line.

\paragraph{No losses ($\alpha_m = 0$ \& $x_{Cm} = 0$):}
This case leads to $\gamma_m = j \beta_m$, and $b_m = -\beta_m^2$. Under these circumstances, and using the SVEA, the CMEs reduce to the traditional form usually described in literature.

\paragraph{Traditional lossy case ($\alpha_m \neq 0$ \& $x_{Cm} = 0$):}
This limit gives $b_m = j \beta_m \gamma_m$. In the SVEA, this result shows that the lossy case can be obtained from the traditional CMEs by substituting $j\beta \to \alpha + j\beta$ but only in the exponents. This is indeed the usual assumption made in nonlinear optics~\cite{Stolen1982}.

%%%%%%%%%%%%%%%%%%%%%%%%%%%%%%%%%%%%%%%%%%%%%%%%%%%%%%%%%%%

\section{Examples}\label{sec:3_Examples}
%In this section we present various examples on the use of Eq.~(\ref{eq:CME}).

\subsection{One Tone}\label{sec:31-One}
\paragraph{$N=1$ ($\omega_1$):}
\begin{figure}[t]
    \centering
    % Top panel
    \begin{overpic}[scale=0.4,trim=0 50 0 0,clip]{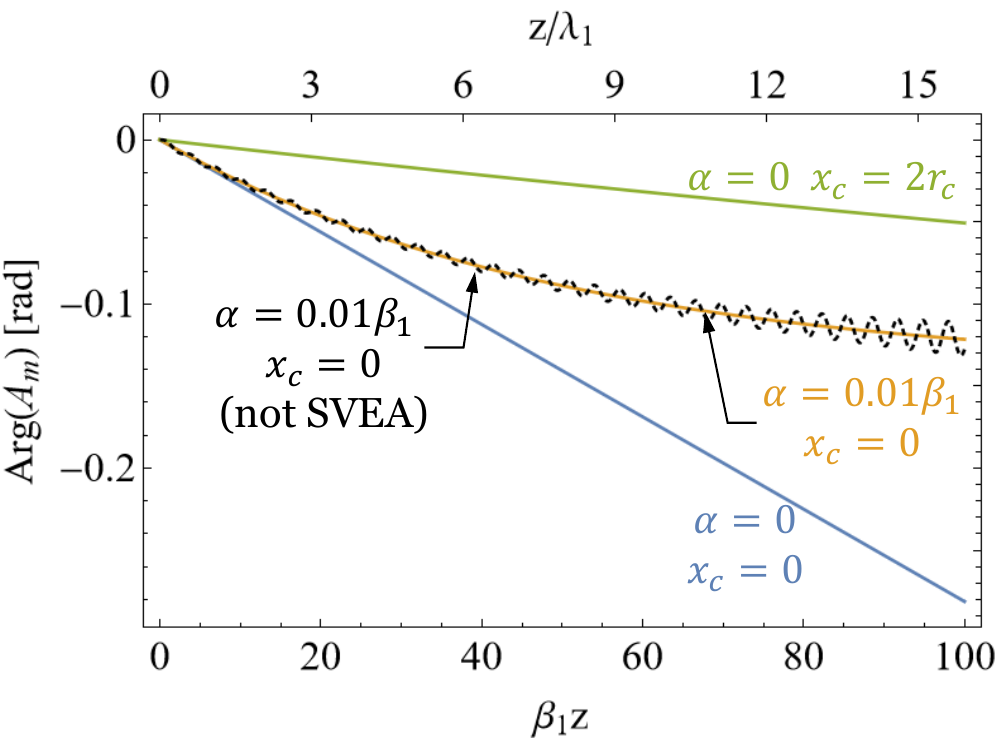}
        \put(-2,55){\footnotesize (a)}
    \end{overpic}\\
    %
    % Bottom panel
    \begin{overpic}[scale=0.4,trim=0 0 0 48,clip]{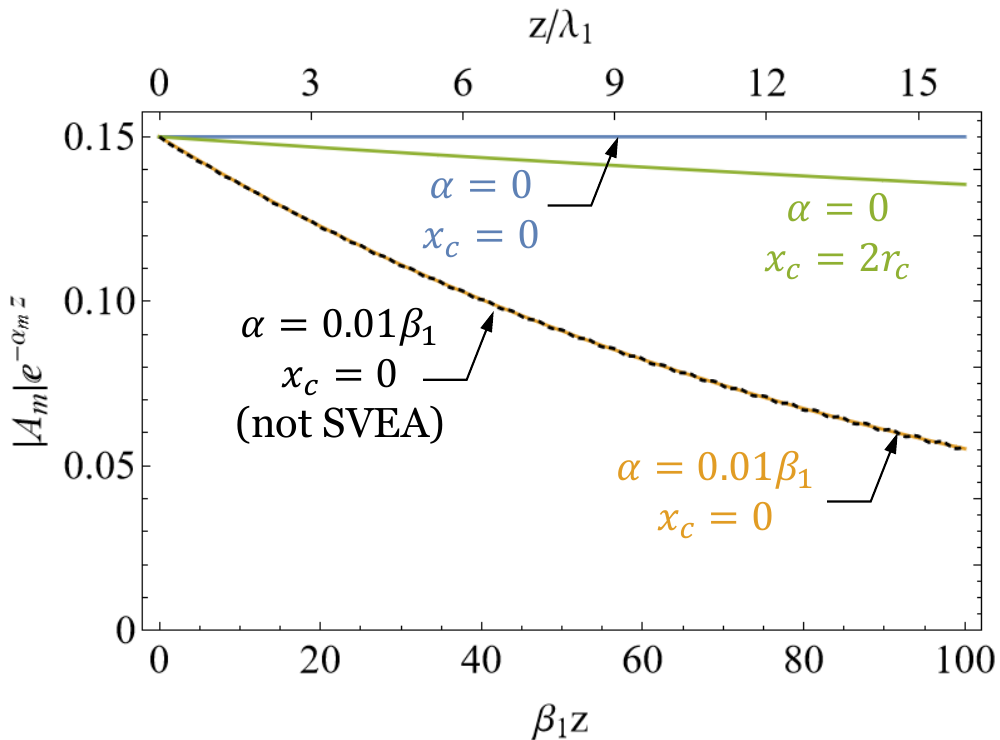}
        \put(-2,58){\footnotesize (b)}
    \end{overpic}
    \caption{Simulation results for one traveling mode, $N=1$. (a) Phase and (b) amplitude. Three cases are considered, a lossless line ($\alpha=x_C=0$) and two with losses, one coming from a complex impedance ($x_c=2r_c$) and one from an attenuation constant ($\alpha=0.01\beta_1$). The simulations were performed within the SVEA for all the cases and without it for the latter case. Adding losses causes an attenuation of the mode and not using the SVEA gives a small spatial oscillation.}
    \label{fig:N1}
\end{figure}
Direct use of (\ref{eq:CME}) gives
\begin{equation*}
    \begin{aligned}
        \mathfrak{A}_1 =\,\, \frac{1}{24} b_1 e^{\gamma_1 z} \Big\{&
        A_1^3 \delta_{\omega_1,3\omega_1} e^{-3\gamma_1 z} \\
        &+ 3 A_1^* A_1^2 \delta_{\omega_1,\omega_1} e^{-(\gamma_1^*+2\gamma_1)z}\\
        &+ 3 A_1^{*2} A_1 \delta_{-\omega_1,\omega_1} e^{-(2\gamma_1^*+\gamma_1)z} \Big\}.
    \end{aligned}
\end{equation*}
Since we are considering one propagating tone and positive frequencies, only one summand is left giving
\[
    \frac{1}{2} A_1'' - \gamma_1 A_1' = \frac{1}{8} b_1 A_1^* A_1^2 e^{-2\alpha_1 z},
\]
where we have also used the definition for $\mathfrak{A}_1$. In the lossless case within the SVEA this equation has a simple analytic solution. By using $A_1 = a_1 e^{\phi_1}$, one obtains that $a_1 = \mathrm{cte}$ and $\phi_1 = -\frac{1}{8}\beta_1 a_1^2 z + \mathrm{cte}$. In other words, only the phase changes linearly with position. Figure~\ref{fig:N1} compares this case with three numerical simulations for specific lossy cases. We can see that the introduction of losses, either as $\alpha \neq 0$ or $x_C \neq 0$, produces a reduction of the amplitude of the mode. Furthermore, the full solution, including $A_1''$, predicts a small spatial oscillation.

\subsection{Harmonic Generation}\label{sec:32_Third}
\paragraph{$N=3$ ($\omega_1$, $\omega_2=2\omega_1$, $\omega_3=3\omega_1$):}
Application of Eq.~(\ref{eq:CME}) and some rewriting gives
\begin{equation*}
\begin{aligned}
\mathfrak{A}_1 &= \frac{b_1}{8}
\Bigl\{
\underbrace{\bigl(e^{\Delta\gamma_1 z}|A_1|^2
+ 2\,e^{\Delta\gamma_2 z}|A_2|^2
+ 2\,e^{\Delta\gamma_3 z}|A_3|^2\bigr) A_1}_{\text{SPM/XPM $\propto A_1$}}\\
&\quad + \underbrace{e^{\Delta\gamma_{1\vert 201}z}A_1^{*2}\,A_3}_{\text{THG back action}}
+ \underbrace{e^{\Delta\gamma_{1\vert 021}z}A_2^2\,A_3^*}_{\text{Source for $A_1$}}
\Bigr\}\\[20pt]
\mathfrak{A}_2 &= \frac{b_2}{8}
\Bigl\{
\underbrace{\bigl(2\,e^{\Delta\gamma_1 z}|A_1|^2
+ e^{\Delta\gamma_2 z}|A_2|^2
+ 2\,e^{\Delta\gamma_3 z}|A_3|^2\bigr) A_2}_{\text{SPM/XPM $\propto A_2$}}\\
&\quad + \underbrace{2\,e^{\Delta\gamma_{2\vert 111}z}A_1\,A_2^*\,A_3}_{\text{SPM/XPM $\propto A^*_2$}}
\Bigr\}\\[20pt]
\mathfrak{A}_3 &= \frac{b_3}{24}
\Bigl\{
\underbrace{e^{\Delta\gamma_{3\vert 300}z}A_1^3
+ 3\,e^{\Delta\gamma_{3\vert 120}z}A_1^*\,A_2^2}_{\text{THG}}\\
&\quad + \underbrace{3\bigl(2\,e^{\Delta\gamma_1 z}|A_1|^2
+ 2\,e^{\Delta\gamma_2 z}|A_2|^2
+ e^{\Delta\gamma_3 z}|A_3|^2\bigr) A_3}_{\text{SPM/XPM $\propto A_3$}}
\Bigr\}
\end{aligned}
\end{equation*}
where $\Delta\gamma_m = -2\,\mathrm{Re}(\gamma_m)$, $\Delta\gamma_{1|201} = \gamma_1 - 2\gamma_1^* - \gamma_3$,
$\Delta\gamma_{1|021} = \gamma_1 - 2\gamma_2 - \gamma_3^*$,
$\Delta\gamma_{2|111} = \gamma_2 - \gamma_2^* - \gamma_1 - \gamma_3$,
$\Delta\gamma_{3|300} = \gamma_3 - 3\gamma_1$,
$\Delta\gamma_{3|120} = \gamma_3 - \gamma_1^* - 2\gamma_2$. As usually discussed in nonlinear optics~\cite{Agrawal2019,Boyd2008,Marhic2007}, we can distinguish several contributions in these equations. First, we have self-phase and cross-phase modulations (SPM \& XPM) that are proportional to the same mode being described by $\mathfrak{A}_m$. Therefore, they do not contribute to spontaneous generation. Since $\mathfrak{A}_2$ only contains SPM and XPM, the second harmonic remains zero if it was not present at $z=0$. It should also be noted that all of the SPM/XPM terms are affected by losses since they are proportional to $e^{\Delta\gamma_{m}}$.
The third equation also contains a third-harmonic generation (THG) term that is not proportional to $A_3$, allowing its spontaneous generation. Finally, the first equation contains two terms proportional to $A_3$. One is the THG back action that represents the energy exchange between the source and the spontaneously-generated third harmonic. The other term is not proportional to $A_1$ representing, then, a source that is only present if $A_2$ is injected at $z=0$. As a consistency check, we have verified that in the lossless limit the resulting equations satisfy the Manley-Rowe relations.

\paragraph{$N=7$ ($\omega_n=n\omega_1$ with $n=1,\ldots, N$):}
\begin{figure}[t]
    \centering
    %Top panel
    \begin{overpic}[scale=0.4,trim=0 46.5 0 0,clip]{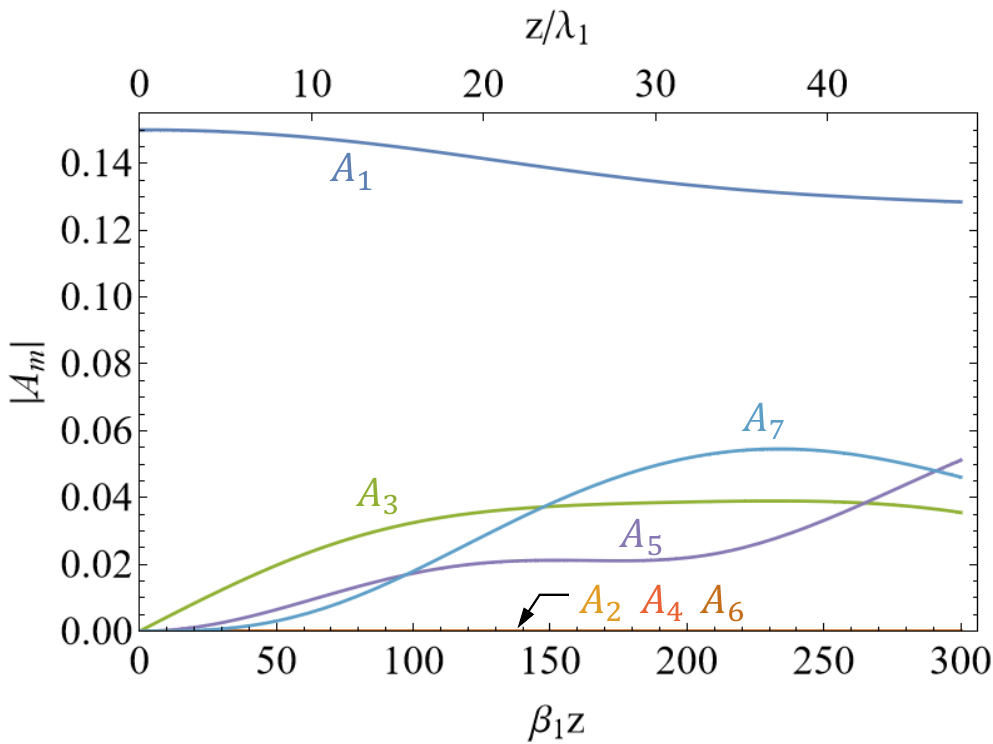}
        \put(-2,52){\footnotesize (a)}
    \end{overpic}\\
    % Center panel
    \begin{overpic}[scale=0.4,trim=0 47 0 54,clip]{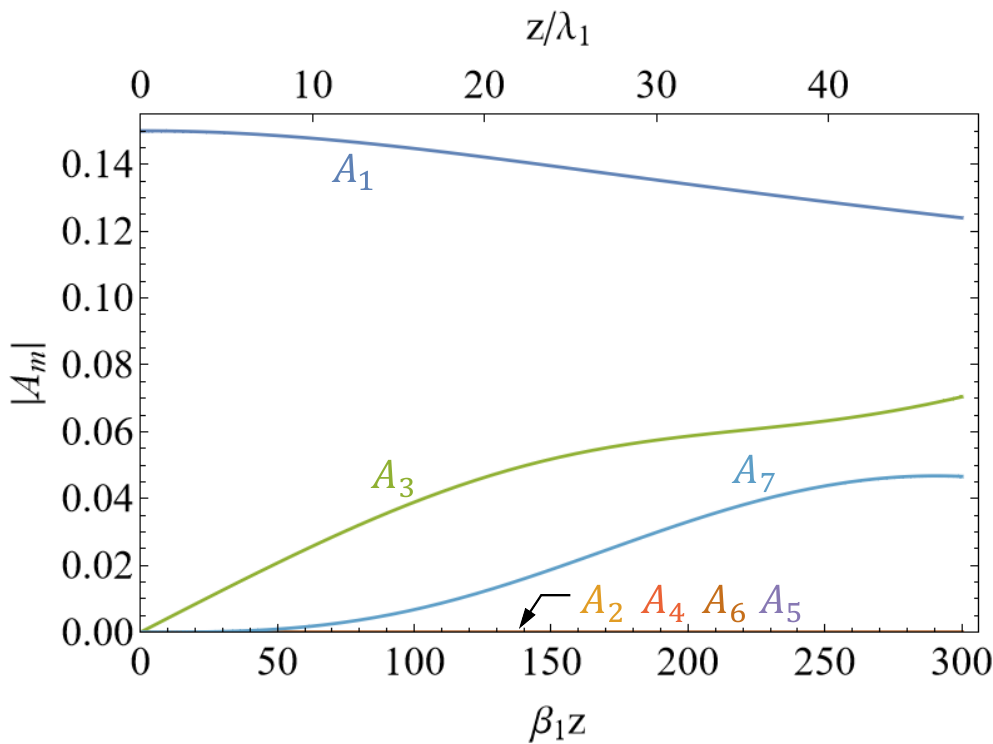}
        \put(-2,50){\footnotesize (b)}
    \end{overpic}\\
    %Bottom panel
    \begin{overpic}[scale=0.4,trim=0 0 0 54,clip]{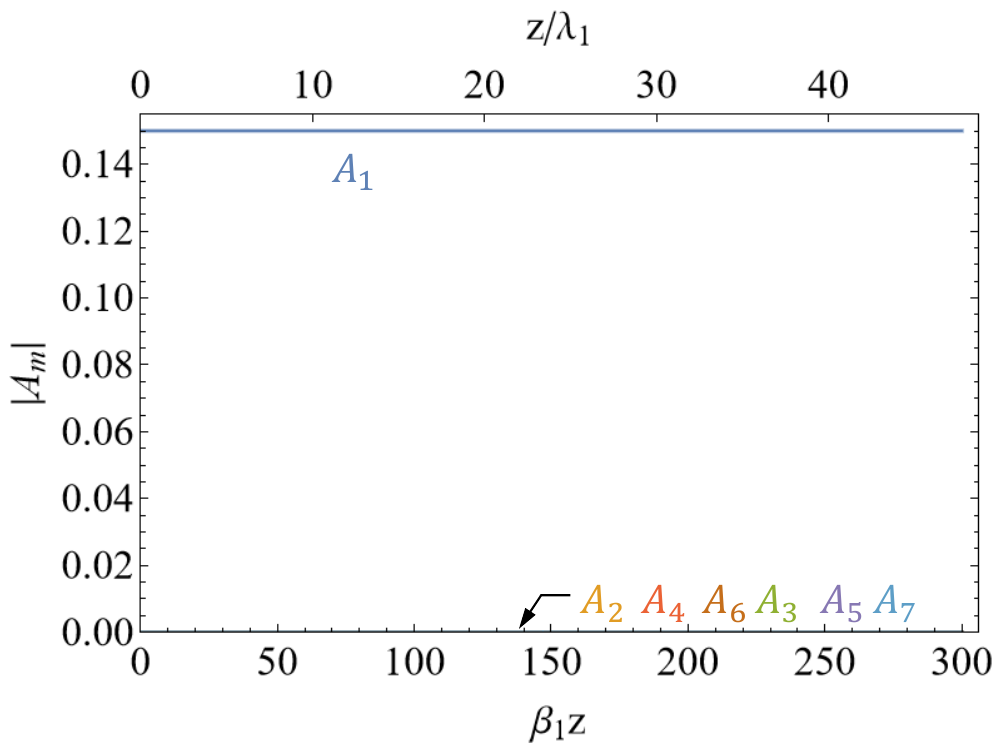}
        \put(-2,60){\footnotesize (c)}
    \end{overpic}
    \caption{Simulation results for harmonic generation with $N=7$ ($\omega_n=n\omega_1$).  For the simulations we set $A_1(0)=0.15$, and assumed linear dispersion, $\beta_n=n\beta_1$, except when they are suppressed, in which case we set $\beta_i\approx 0$. (a)~All tones are allowed to travel. Only odd harmonics are spontaneously excited. (b)~Fifth harmonic is suppressed resulting in increased THG, although the seventh harmonic still propagates. (c)~Third harmonic is suppressed resulting in none of the other harmonics being excited.}
    \label{fig:N7}
\end{figure}
In order to better illustrate the use of (\ref{eq:CME}), we have performed three simulations where one tone and six harmonics are included in the model. For the sake of clarity, we have taken the lossless case within the SVEA and the results are presented in Fig.~\ref{fig:N7}. First, panel~(a), we considered a transmission line with linear dispersion, i.e. $\beta_n = n\beta_1$. It can be seen that, as expected, only odd harmonics are generated. Second, panel~(b), we simulated the suppression of the fifth harmonic by setting $\beta_5 \approx 0$. We can see that it increases the generation of the third harmonic but it does not suppress higher-order modes. The last simulation, panel~(c), represents the suppression of the third harmonic ($\beta_3 \approx 0$), which leads to the suppression of all other harmonics. In principle, this result suggests that in parametric amplifiers, the third harmonic is the most important one to be suppressed.

Additionally, we have studied Eq.~(\ref{eq:CME}) with $N=4$ ($\omega_1$, $\omega_2=3\omega_1$, $\omega_3=5\omega_1$, $\omega_4=7\omega_1$). We have verified that this situation is equivalent to the case studied above when the initial amplitudes of the even harmonics are kept equal to zero.

\subsection{Degenerate Four-Wave Mixing (FWM) with \& without Higher-Order Effects}\label{sec:33_FWM}
\paragraph{$N=3$ ($\omega_1$, $\omega_2$, $\omega_3=2\omega_1-\omega_2$):}
This is the traditional FWM case where $\omega_1 \to \mathrm{pump}$, $\omega_2 \to \mathrm{signal}$, and $\omega_3 \to \mathrm{idler}$. The resulting system of equations is
\[
\begin{aligned}
    \mathfrak{A}_1 = \frac{1}{8} b_1 \Big[&
    \big( e^{\Delta\gamma_{1} z} |A_1|^2
    + 2 e^{\Delta\gamma_{2} z} |A_2|^2
    + 2 e^{\Delta\gamma_{3} z} |A_3|^2 \big) A_1 \\
    &+ 2 e^{(\Delta\gamma_1-\Delta\gamma)z} A_1^* A_2 A_3
    \Big]
    \\[6pt]
    \mathfrak{A}_2 = \frac{1}{8} b_2 \Big[&
    \big( 2 e^{\Delta\gamma_{1} z} |A_1|^2
    + e^{\Delta\gamma_{2} z} |A_2|^2
    + 2 e^{\Delta\gamma_{3} z} |A_3|^2 \big) A_2 \\
    &+ e^{(\Delta\gamma_{3}+\Delta\gamma)z}  A_1^2 A_3^*
    \Big]
    \\[6pt]
    \mathfrak{A}_3 = \frac{1}{8} b_3 \Big[&
    \big( 2 e^{\Delta\gamma_{1} z} |A_1|^2
    + 2 e^{\Delta\gamma_{2} z} |A_2|^2
    + e^{\Delta\gamma_{3} z} |A_3|^2 \big) A_3 \\
    &+ e^{(\Delta\gamma_{2}+\Delta\gamma)z}  A_1^2 A_2^*
    \Big],
    \end{aligned}
\]
where $\Delta\gamma_{m} = -2\operatorname{Re}(\gamma_m)$ and $\Delta\gamma = \gamma_3 + \gamma_2 - 2\gamma_1$. Here we can note that all terms contain $e^{\Delta\gamma_m}$ and, consequently, are affected by losses. It can also be demonstrated that the system reduces to the traditional one in the lossless case.

\paragraph{$N=9$:}
\begin{figure}[!t]
    \centering
    \includegraphics[scale=0.4]{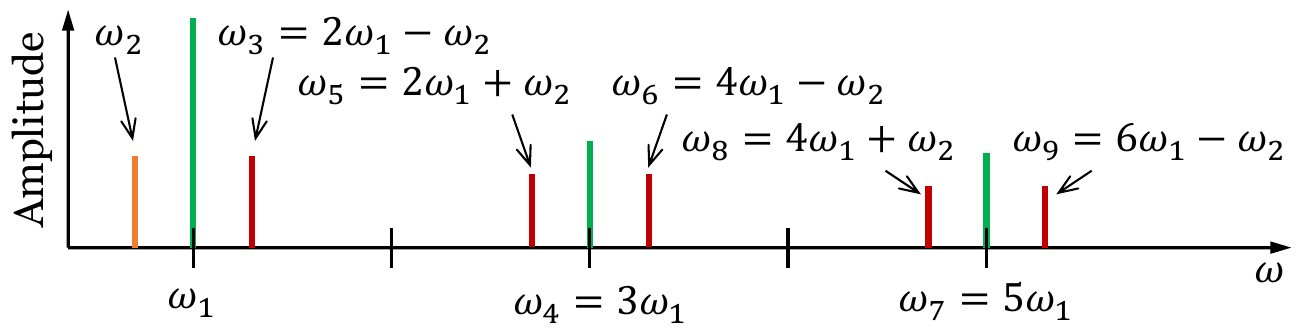}
    \caption{Cartoon of the degenerate FWM with higher-order effects. Specifically, two harmonics and their idlers are allowed to propagate. Nine modes are produced in this process.}
    \label{fig:N9_scheme}
\end{figure}
This case corresponds to the one depicted in Fig.~\ref{fig:N9_scheme} where we consider pump and two harmonics ($\omega_1$, $\omega_4$, $\omega_7$), signal ($\omega_2$), and idlers ($\omega_3$, $\omega_5$, $\omega_6$, $\omega_8$, $\omega_9$). To illustrate the effect of harmonics and high-order idlers in the parametric gain of the signal, we have performed various simulations presented in Fig.~\ref{fig:N9}. Again, in order to simplify the discussion, we have considered the lossless case within the SVEA. Furthermore, we have performed the simulations with perfect matching between the pump, signal and first idler, i.e., $\Delta\beta = -|A_1(0)|^2 \Delta\overline{\beta}$, where $\Delta\beta = \beta_3 + \beta_2 - 2\beta_1$ and $\Delta\overline{\beta} = \frac{1}{4} \left( \beta_3 + \beta_2 -\beta_1 \right)$. Notice that this condition is more general and reduces to the standard form usually presented in the literature where it is assumed that $\beta_1 \approx \beta_2$~\cite{Eom2012,Hansryd2002}. For the rest of the waves we have considered linear dispersion unless specified otherwise.

\begin{figure}[t]
    \centering
    % Top panel
    \begin{overpic}[scale=0.4,trim=0 51 0 0,clip]{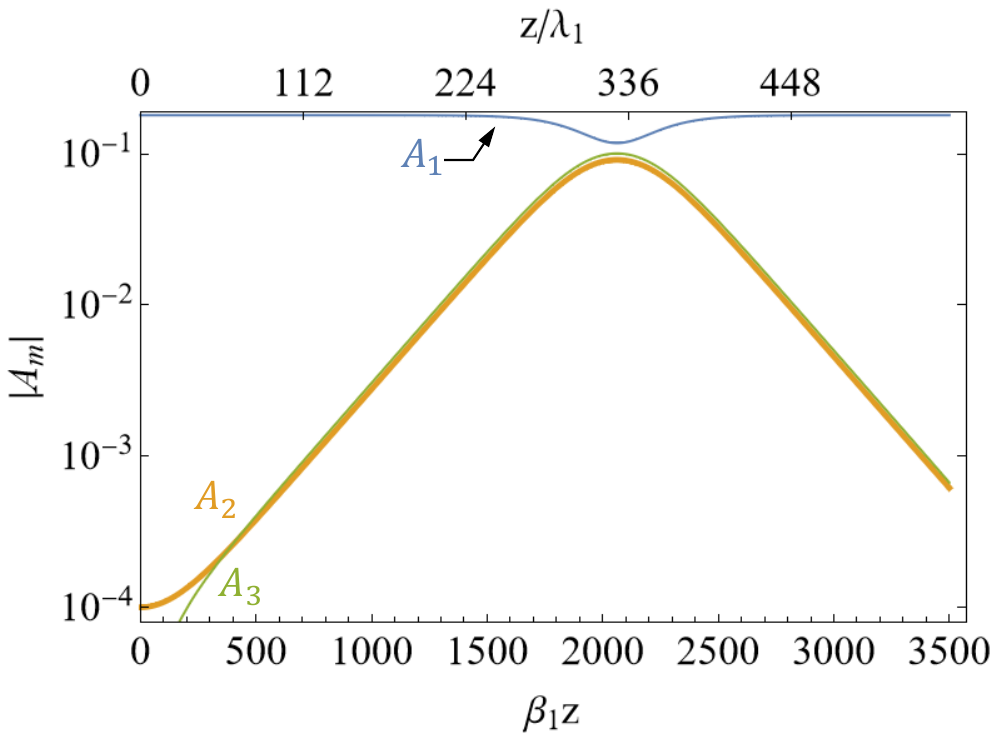}
        \put(-2,52){\footnotesize (a)}
    \end{overpic}\\
    % Center panel
    \begin{overpic}[scale=0.4,trim=0 52 0 52,clip]{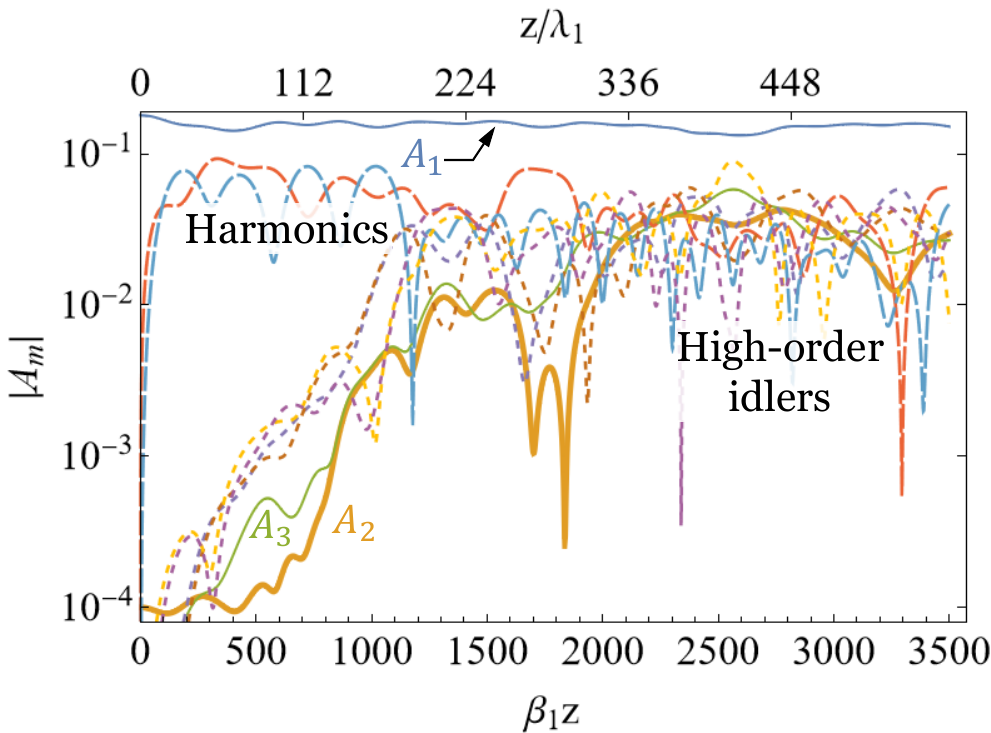}
        \put(-2,50){\footnotesize (b)}
    \end{overpic}\\
    % Bottom panel
    \begin{overpic}[scale=0.4,trim=0 0 0 52,clip]{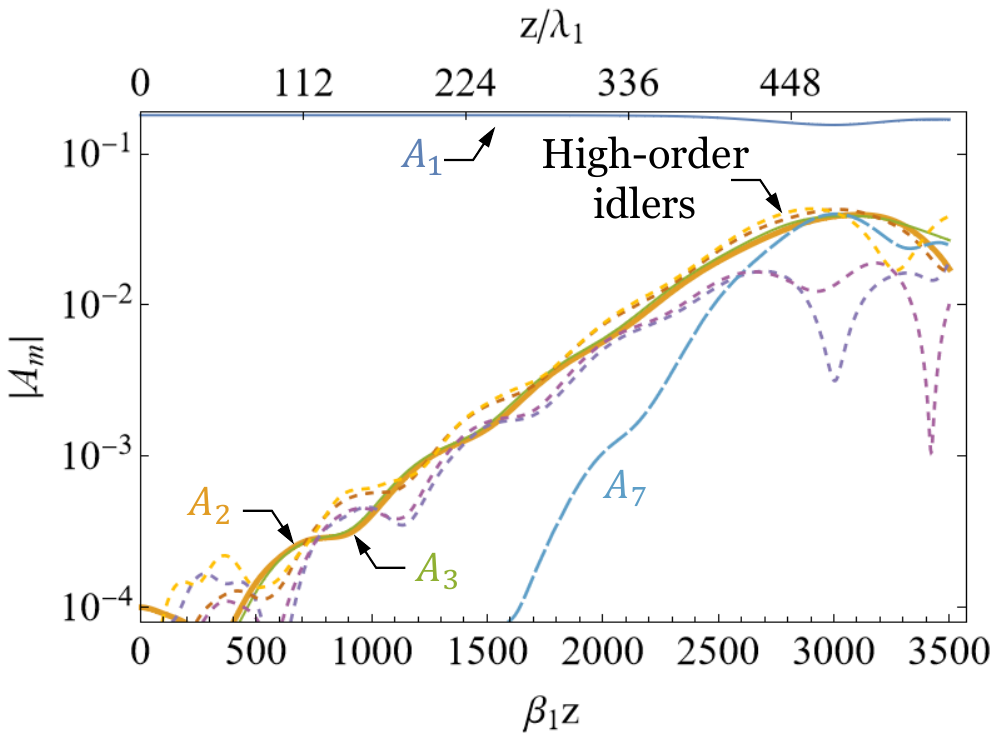}
        \put(-2,60){\footnotesize (c)}
    \end{overpic}
    \caption{Simulation results for FWM (pump, $\omega_1$, signal, $\omega_2$, idler, $\omega_3$) with two harmonics ($\omega_4$, $\omega_7$) and high-order idlers ($\omega_5$, $\omega_6$, $\omega_8$, $\omega_9$), $N=9$, corresponding to the situation depicted in Fig.~\ref{fig:N9_scheme}. For the simulations we set $A_1(0)=0.18$, $\beta_2=0.9\beta_1$, $\beta_3=1.092\beta_1$, ensuring perfect matching as described in the text. For the rest of the modes we assumed a linear dispersion, except when they are suppressed, in which case we set $\beta_i\approx 0$. (a)~Ideal FWM case obtained by suppressing all harmonics and high-order idlers. Notice the exponential gain achieved by perfect matching. (b)~All nine modes are allowed to propagate. (c)~The third harmonic ($\omega_4=3\omega_1$) is suppressed.}
    \label{fig:N9}
\end{figure}
Panel (a) considers the ideal degenerate FWM, meaning that only pump, signal and first idler are allowed to propagate. We achieve this situation by setting all other propagation constants close to zero. The perfect-matching condition is revealed by the exponential gain of the signal until saturation occurs.

In the simulation depicted in panel~(b), all modes are allowed to propagate. We can see that, despite the fact that all of them interact, a signal gain still occurs.

In panel~(c), the simulation allows all the waves but the third harmonic to propagate ($\omega_4=3\omega_1$, $\beta_4 \approx 0$).  As anticipated in the previous subsection, the fifth harmonic is initially suppressed, but eventually excited. Its excitation and that of the high-order idlers limit the signal gain resulting in the need of a longer line to obtain a similar gain to that shown in panel (a). The signal gain above $\beta_1 z \approx 3000$ can be improved if the matching condition is changed slightly or if the fifth harmonic is also suppressed. However, more importantly, the ideal situation of panel (a) is recovered if the idlers around the third harmonic ($\omega_5$ and $\omega_6$) are not allowed to propagate either. These simulations confirm that neglecting harmonics and high-order idlers can affect the achievable gain~\cite{Klimovich2024}.
%The effect of the idlers on the signal gain has also been studied in~\cite{Klimovich2024}.

%%%%%%%%%%%%%%%%%%%%%%%%%%%%%%%%%%%%%%%%%%%%%%%%%%%%%%%%%%%

\section{Experiment \& Methodology}\label{sec:4_Exp}
In order to demonstrate the validity of the CMEs and the approach presented here, we have studied multiple harmonic generation in a superconducting transmission line. To achieve a parameter-free validation we have also determined $I'_*$ independently by studying the dependence of the transmission of the sample, $S_{21}$, on applied DC current. The value found for $I'_*$ justifies our selection of harmonic generation over FWM for validation since it requires lower pump currents. Moreover, we complement these measurements with a determination of the critical current from resistance-current measurements using a sense resistor.

\subsection{Sample}\label{sec:41_sample}

\begin{table}[!t]
  \caption{Material Properties}
  \label{tab:materials}
  \centering
  \begin{tabular}{ccccc}
    \toprule
    Si & \multicolumn{3}{c}{NbTiN} \\
    \cmidrule(lr){1-1}\cmidrule(lr){2-4}
    $\epsilon_r$ & $t$ & $T_c$ & $\rho_n$ \\
    {--} & \si{\nano\meter} & \si{\kelvin} & \si{\micro\ohm\centi\meter} \\
    \midrule
    11.8 & 60 & 14.7 & 132 \\
    \bottomrule
  \end{tabular}
\end{table}
\begin{figure}[t]
    \centering
    \includegraphics[scale=0.4]{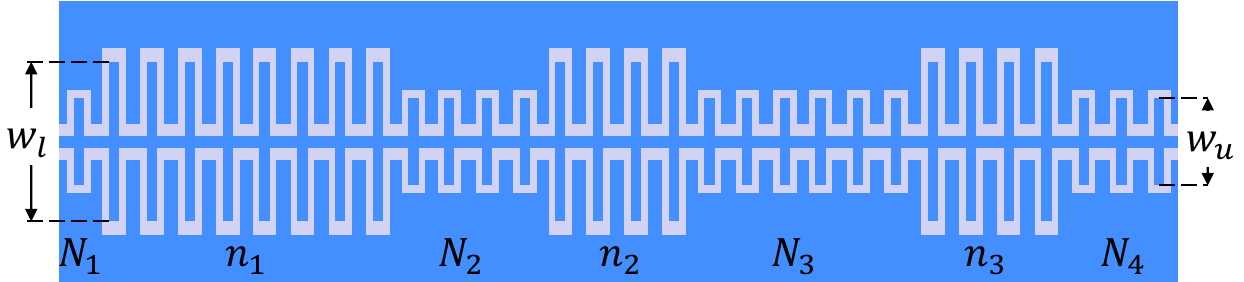}
    \caption{Schematics of the unit cell of the CPW DE line used in this work. Three loads, $n_i$, are intercalated along a central line. The characteristics of the resulting sections are given in Table~\ref{tab:UC_par}. The CPW uses a NbTiN layer deposited over a Si substrate. The nominal material properties are given in Table~\ref{tab:materials}.}
    \label{fig:1b_scheme}
\end{figure}
\begin{table}[t]
	\caption{Parameters of Each Section of the Unit Cell.}
	\label{tab:UC_par}
	\centering
	\footnotesize
	\setlength{\tabcolsep}{4pt}
	\renewcommand{\arraystretch}{1.1}
	\begin{tabular}{c c c c c c c c}
		\toprule
			& $N_1$ & $N_2$ & $N_3$ & $N_4$ & $n_1$ & $n_2$ & $n_3$ \\
		\cmidrule(lr){2-5} \cmidrule(lr){6-8}
			\# Stubs & 54 & 79 & 89 & 27 & 40 & 20 & 20 \\
			$w\,(\mu\mathrm{m})$   & \multicolumn{4}{c}{23}    & \multicolumn{3}{c}{38} \\
			$Z_0\,(\Omega)$        & \multicolumn{4}{c}{66}    & \multicolumn{3}{c}{55} \\
			$v/c$                  & \multicolumn{4}{c}{0.092} & \multicolumn{3}{c}{0.073} \\
			$\alpha_K$                  & \multicolumn{4}{c}{0.62} & \multicolumn{3}{c}{0.58} \\
		\bottomrule
	\end{tabular}
\end{table}
\begin{figure}[t]
    \centering
    % Top panel
    \begin{overpic}[scale=0.40,trim=0 50 0 0,clip]{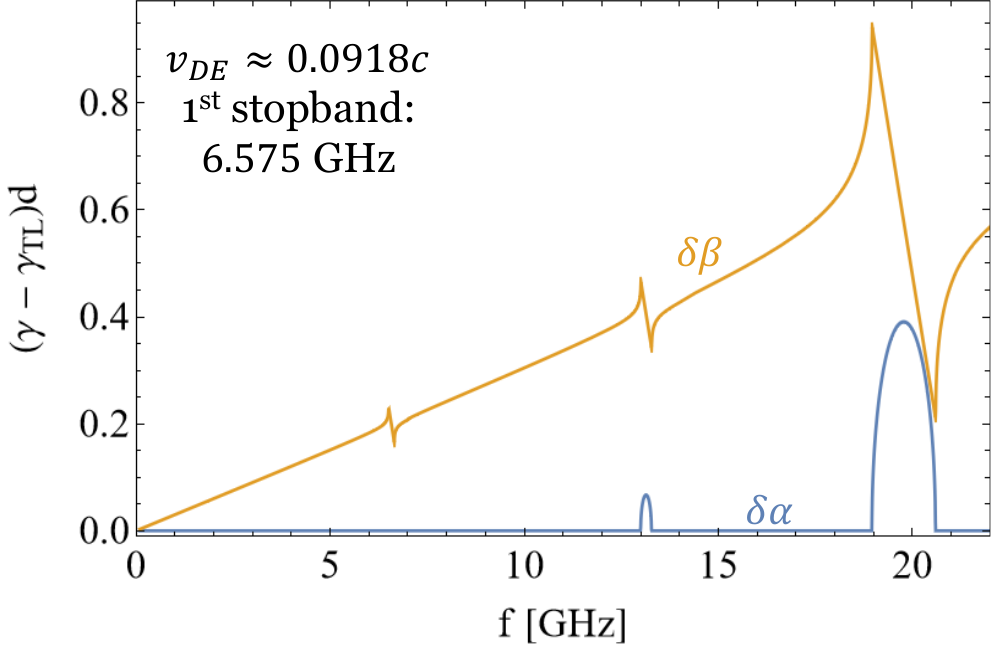}
        \put(0,52){\footnotesize (a)}
    \end{overpic}\\
    % Bottom panel
    \begin{overpic}[scale=0.4,trim=0 0 0 0,clip]{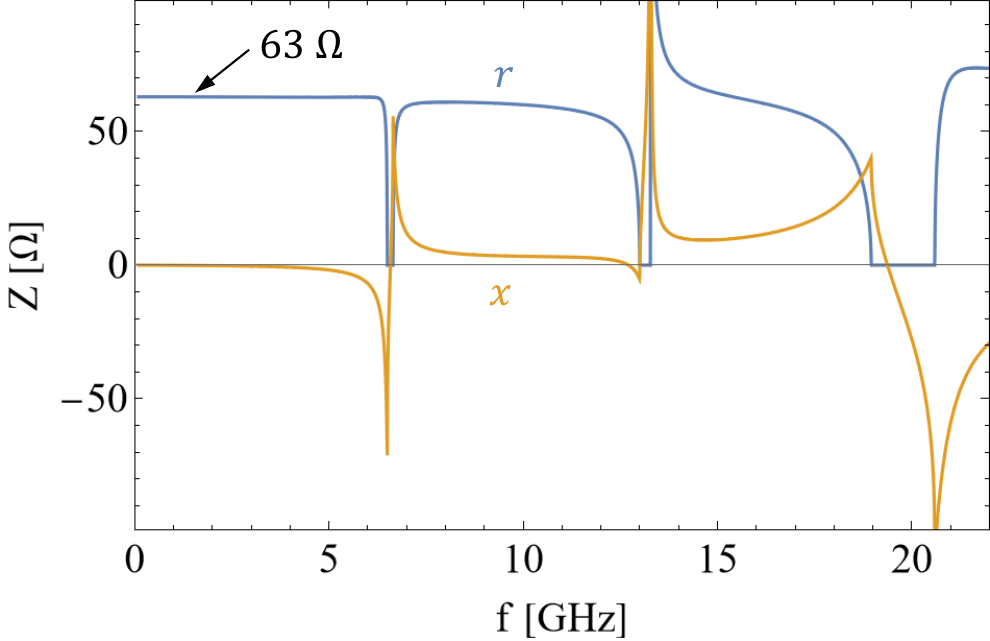}
        \put(0,60){\footnotesize (b)}
    \end{overpic}
    \caption{Simulated properties of the sample used in this work. (a)~Propagation constant of the sample, $\gamma$, minus the propagation constant of the central line, $\gamma_{\mathrm{TL}}$. This difference is normalized by $d=\SI{1974}{\micro\meter}$, the length of the unit cell. (b)~Characteristic impedance. Notice the strong frequency dependence, especially around the stopbands.}
    \label{fig:1b_gz}
\end{figure}
The sample used in this study is a CPW capacitively-loaded artificial superconducting transmission line from the same batch used in~\cite{Mena2024} and packaged in the same way. The superconductor is NbTiN deposited over a Si substrate, whose nominal properties are given in Table~\ref{tab:materials}. It has an engineered dispersion obtained by repeating 48 times the unit cell depicted in Fig.~\ref{fig:1b_scheme} with the number of capacitive stubs given in Table~\ref{tab:UC_par}. The central strip width and the gap to the ground plane are the same and equal to \SI{1.5}{\micro\meter}. The characteristic impedance and phase velocity of each section, presented also in Table~\ref{tab:UC_par}, were obtained by simulation, using the methods described in~\cite{Mena2024}. Those values were used in turn to determine the properties of the entire DE transmission line (Fig.~\ref{fig:1b_gz}).

\subsection{Experimental Setups}\label{sec:42_setups}
\begin{figure}[t]
    \centering
    \begin{overpic}[scale=0.4]{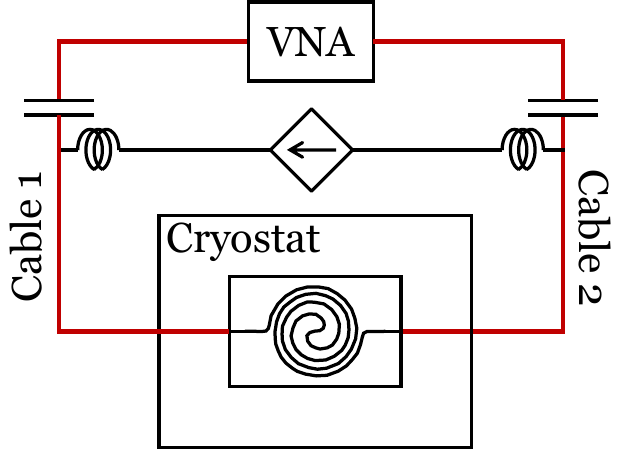}
        \put(48,-10){\footnotesize (a)}
    \end{overpic}
    \hfill
    \begin{overpic}[scale=0.4]{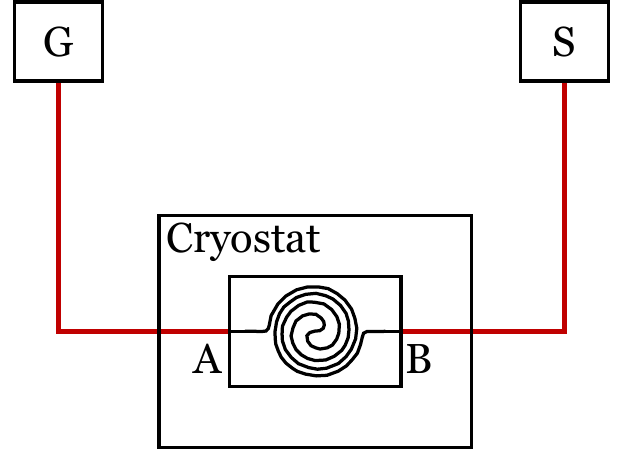}
        \put(48,-10){\footnotesize (b)}
    \end{overpic}
    \vspace{0.8em}
    \caption{Experimental setups. (a)~Measurement of the dependence of $S_{21}$ on applied DC current. The sample is kept in a cryostat, and it is connected to a VNA and a current source simultaneously using external bias tees. (b)~A similar configuration is used to measure harmonic generation. The sample is  connected at points~A and~B, via coaxial cables, to a signal generator, G, and a spectrum analyzer, S. No additional amplifiers were used to avoid their saturation or own harmonic generation. The maximum operating frequency of the VNA and spectrum analyzer is \SI{20}{\giga\hertz}.}
    \label{fig:Exp_setup}
\end{figure}
As described in Fig.~\ref{fig:Exp_setup}, with the sample inside a cryostat, we measured $S_{21}$ under different DC currents, panel~(a), and harmonic generation, panel~(b). For the latter, in order to determine the actual incident and exiting powers, $P_A$ and $P_B$, we performed an additional measurement. We removed the sample, terminated points~A and~B with shorts, and measured at cryogenic temperatures the resulting reflections. Assuming perfect matching of the cabling to the VNA, the reflections give two times the losses of each path. With the sample remaining in the cryostat, we also obtained a resistance-current plot using a two-point measurement with a sense resistor.

\subsection{Model for $I_*$ Determination}\label{sec:43_S21}
For modeling this experiment, following panel (a) of Fig.~\ref{fig:Exp_setup}, we start by constructing a transmission-matrix model of the signal path, $T_{c1}.T_{DE}.T_{c2}$, i.e., the DE sample between two lossy cables. We consider the case where we are well away from the stopbands so the sample is lossless resulting in $\gamma=j\beta$ and $Z=r$. Then, the resulting ABCD matrix is converted to an $S$ matrix assuming that the cables are perfectly matched to the VNA. The transmission term of such a matrix  is calculated to be
\begin{equation}\label{eq:S21}
S_{21}=\frac{2j\,e^{-(l_{c1}+l_{c2})\gamma_c}\,Z Z_c}{2j Z Z_c \cos(\beta l)-(Z^2+Z_c^2)\sin(\beta l)},
\end{equation}
where $Z_c$, $\gamma_c$, and $l_{ci}$ are the impedance, propagation constant, and lengths of the cables, and $l$ is the total length of the sample. Now, when we apply a DC current to the system, due to the nonlinear inductance Eq.~(\ref{eq:Ltot}) and the fact that we are away from the stopbands, the impedance and propagation constant of the sample change, respectively, according to $Z_i = g Z_o$ and $\beta_i = g \beta_o$, where
\[
g= \big[ 1+ (I_{DC}/I'_*)^2\big]^{1/2}.
\]
In these expressions, $Z_o$ and $\beta_o$ are the values when $I_{DC}=0$. Finally, using Eq.~(\ref{eq:S21}), we calculate the ratio of $S_{21}$ with and without an applied current giving
\begin{equation}\label{eq:S21ratio}
    \frac{S_{21}(I_{DC})}{S_{21}(0)}
    =
    \frac{Z_i}{Z_o}
    %\,
    \frac{
    2j Z_o Z_c \cos(\beta_o l) - (Z_o^2 + Z_c^2)\sin(\beta_o l)
    }{
    2j Z_i Z_c \cos(\beta_i l) - (Z_i^2 + Z_c^2)\sin(\beta_i l)
    },
\end{equation}
which removes the need for an independent characterization of the cables.

Equation~(\ref{eq:S21ratio}) can be used to fit experimental data. Since $\beta_o$ and $Z_o$ can be taken from Fig.~\ref{fig:1b_gz}, and $Z_c = \SI{50}{\ohm}$, $I'_*$ becomes the only free parameter of the model.

\subsection{Model for Harmonic Generation}\label{sec:44_HG}
\begin{figure}[t]
    \centering
    \includegraphics[scale=0.4]{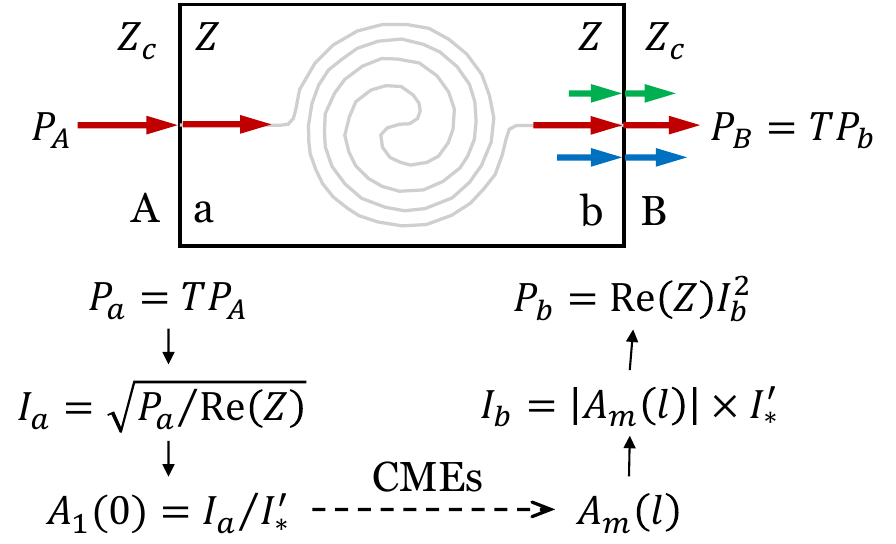}
    \caption{Model for the study of harmonic generation. A fraction of the arriving input power, $P_A$, enters the sample, $P_a$, via the transmission coefficient $T=4\operatorname{Re}(Z_c Z^*)/|Z_c+Z|^2$ and generates the rms current $I_a$. This current is expressed as a normalized amplitude, $A_1(0)$, using $I'_*$. Through the CMEs, the resulting normalized amplitudes $A_m(l)$ at the end of the transmission line are calculated. They are converted back to current, $I_b$, and power inside the sample, $P_b$. Finally, the output powers generated in the coaxial output, $P_B$, are calculated. The impedance of the coaxial feeds was taken as $Z_c=\SI{50}{\ohm}$ while the impedance $Z$ of the sample corresponds to that of the different harmonics as given by Fig.~\ref{fig:1b_gz}.}
    \label{fig:Model}
\end{figure}
To analyze the harmonic-generation experiment, we prepared the model detailed in Fig.~\ref{fig:Model}. Its input is the transmitted power of the injected tone into the sample, $P_a$. The power is converted to rms current and normalized by $I'_*$, giving the initial condition for use in the CMEs. The reversed process gives the output powers $P_B$ of the different harmonics.

For the CMEs, we took the lossless case with $N=4$ ($\omega_1$, $\omega_2=3\omega_1$, $\omega_3=5\omega_1$, $\omega_4=7\omega_1$) within the SVEA approximation. We neglected losses for two reasons. First, the measurements were performed away from the stopbands where material losses are small and, second, within the experimental uncertainty of our measurements, their effect on harmonic generation could not be distinguished from the lossless case. The complex propagation constants and impedances required by the model were those given in Fig.~\ref{fig:1b_gz}. In this manner, as with the previous model, the only free parameter is $I'_*$.

%%%%%%%%%%%%%%%%%%%%%%%%%%%%%%%%%%%%%%%%%%%%%%%%%%%%%%%%%%%%%%%%%%%%%%

\section{Results and Discussion}\label{sec:5_R&D}

\subsection{Resistance-Current Measurements}\label{sec:50_RI}
\begin{figure}[t]
    \centering
    \includegraphics[scale=0.4]{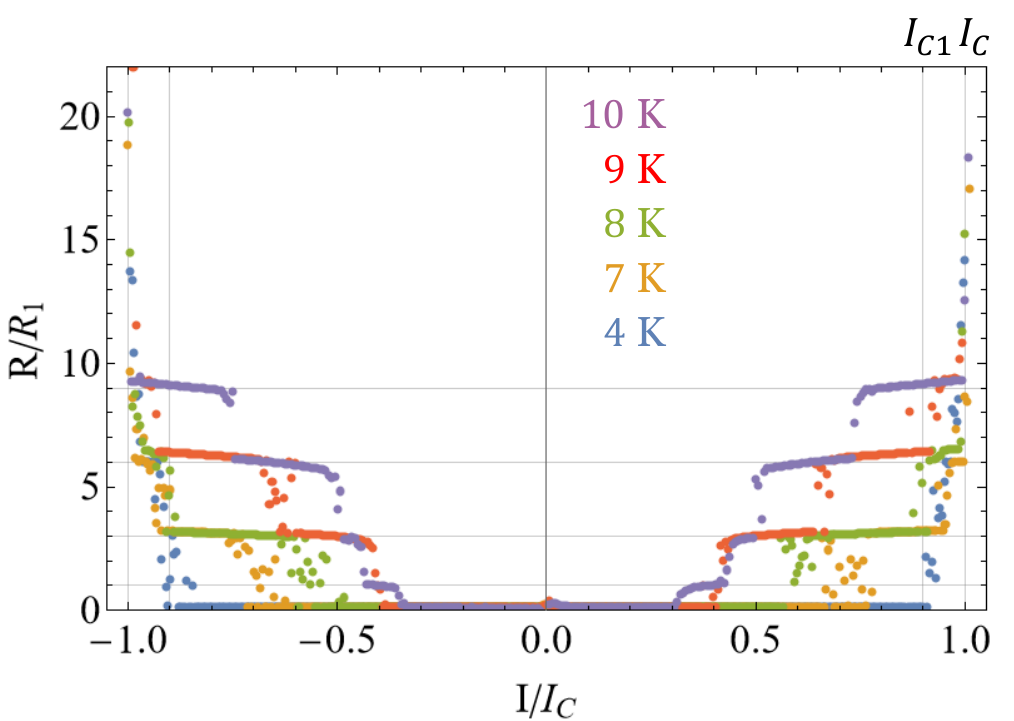}
    \caption{Normalized resistance-current measurement at different temperatures. Several steps, attributed to the presence of dissipative regions, and their evolution with temperature are clearly seen. The resistance is normalized by $R_1 =\SI{40}{\ohm}$, the resistance reached by the first step at \SI{10}{\kelvin}. This temperature corresponds to the highest temperature with a well defined non-resistive region. At $\SI{4}{\kelvin}$, the onset of the first step is located at $I_{C1} = \SI{1.8}{\milli\ampere}$, while $I_C=\SI{2}{\milli\ampere}$.}
    \label{fig:RI}
\end{figure}
In Fig.~\ref{fig:RI}, we present the results of the resistance-current measurements, in normalized form, at different temperatures. The main common feature for all the curves is the presence of steps whose resistance values are independent of temperature. In contrast, the current onset of each step has a strong temperature dependence. Relevant for the measurement presented in the next sections is the first plateau with an onset $I_{C1}=0.9I_C$ at \SI{4}{\kelvin}.

We notice that the steps reach a resistance that is only a small fraction of the resistance in the normal state, $R_n\approx \SI{1.5}{\mega\ohm}$, with a value $R_1=\SI{40}{\ohm}$ for the first step at \SI{10}{\kelvin}. These steps collapse to discrete multiples of $R_1$ indicating highly localized dissipation. The systematic shift of the plateaus towards higher currents as the temperature is lowered indicates that they are more difficult to form as the superconducting state becomes more robust. Furthermore, the nearly temperature-independent resistance values suggest that these states correspond to a small number of preferred dissipative configurations associated with the device geometry~\cite{Skocpol1974}.

We also observe that the transition from one step to the other is not a well defined process but accompanied by fluctuations, signaling metastable states between successive dissipative configurations. This situation is further supported by the fact that the exact $R$-$I$ characteristics depend on the current sweep protocol both in direction and current step.

\subsection{Independent Determination of $I_*$}\label{sec:51_S21}
\begin{figure}[t]
    \centering
    \includegraphics[scale=0.4]{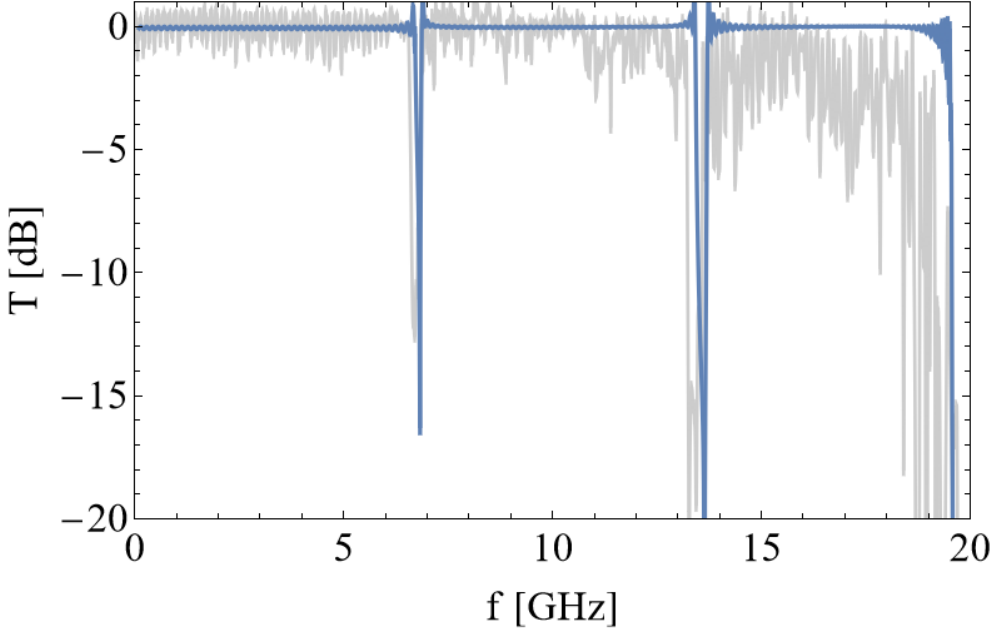}
    \caption{Normalized transmission at \SI{4}{\kelvin} without applied DC current. Measured (gray) and simulated (blue). No fit was attempted.}
    \label{fig:S21}
\end{figure}
We start by showing the measured transmission without applied current at \SI{4}{\kelvin} in Fig.~\ref{fig:S21}. The figure also shows the simulated transmission using the properties of the line presented in Fig.~\ref{fig:1b_gz}. An excellent agreement between measurement and simulation is evident, with the stopbands differing only by a few hundred \si{\mega\hertz}. The small difference can easily be explained by reasonable deviations from the nominal values of the material properties given in Table~\ref{tab:materials}. Moreover, we also notice that the actual stopbands are broadened with respect to the simulation, likely due to either the finite number of unit cells or unmodeled losses in the line.

\begin{figure}[t]
    \centering
    % Top panel
    \begin{overpic}[scale=0.4,trim=0 53 0 0,clip]{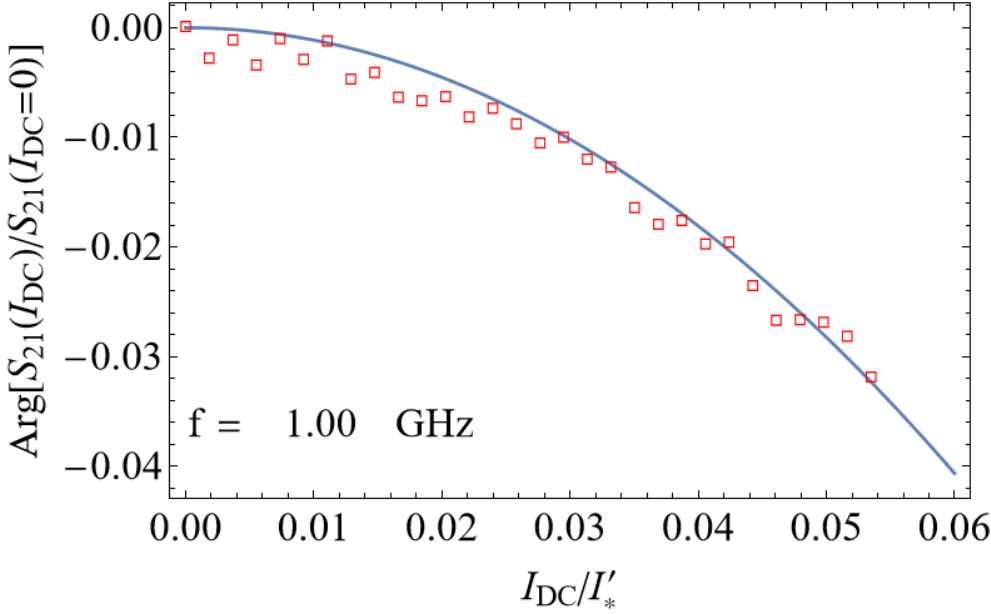}
        \put(91,45){\footnotesize (a)}
    \end{overpic}\\
    % Center panel
    \begin{overpic}[scale=0.4,trim=0 52 0 0,clip]{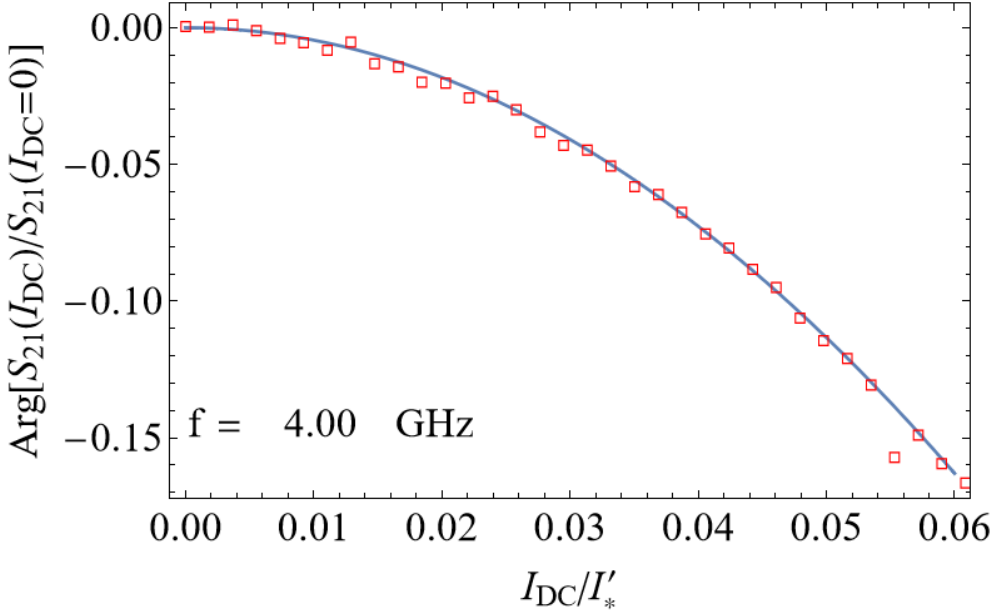}
        \put(91,45){\footnotesize (b)}
    \end{overpic}\\
    % Bottom panel
    \begin{overpic}[scale=0.4,trim=0 0 0 0,clip]{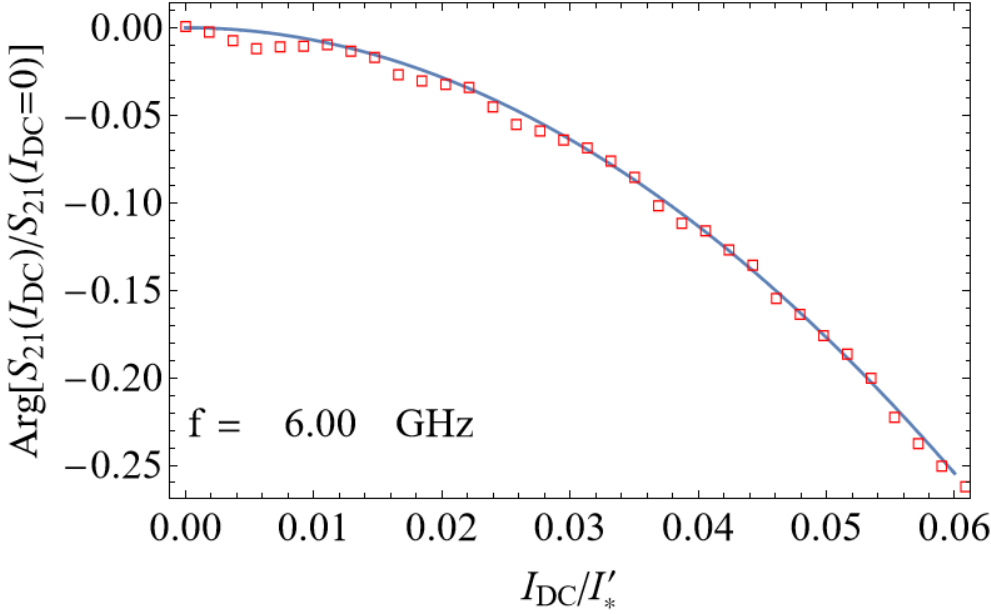}
        \put(91,56){\footnotesize (c)}
    \end{overpic}
    \caption{Examples of the relative phase variation of $S_{21}$ with applied DC current. (a)~\SI{1}{\giga\hertz}. (b)~\SI{4}{\giga\hertz}. (c)~\SI{6}{\giga\hertz}. Empty squares are measured data. Lines are plots of Eq.~(\ref{eq:S21ratio}) using $I'_*=\SI{27}{\milli\ampere}$, the global best-fit parameter at 551 frequency points, as described in the text. }
    \label{fig:S21_exp}
\end{figure}
Next, Fig.~\ref{fig:S21_exp} shows the variation of $S_{21}$ with applied DC current. We performed a nonlinear least-squares fit of Eq.~(\ref{eq:S21ratio}) to the  measured $S_{21}$ variation using 551 frequency points from \SIrange{0.5}{6}{\giga\hertz}. As the best-fit parameter we obtained $I'_* = \SI{27.116(7)}{\milli\ampere}$. Since $\alpha_K \approx 0.6$ (see Table~\ref{tab:UC_par}), we obtain $I_*\approx \SI{21}{\milli\ampere}$. Surprisingly, this nonlinear parameter does not scale with the measured $I_C$ but with the much larger depairing current as calculated from the Ginzburg-Landau theory~\cite{Tinkham1996}. 

From the measured values of the NbTiN film used to fabricate the sample (Table~\ref{tab:materials}), we obtain that its penetration depth is $\lambda \approx \SI{315}{\nano\meter}$~\cite{Zmuidzinas2012}. Moreover, as reported for similar films~\cite{Yu2002,Lee2024}, we can take the coherence length as $\xi\approx\SI{4}{\nano\meter}$. These two values allow us to calculate the depairing density current giving $J_{\mathrm{dep}} \approx \SI{2.54e11}{\ampere\per\metre\squared}$. Now, if we consider the central strip of the DE transmission line, which has a cross section of $\SI{1.5}{\micro\meter}\times\SI{60}{\nano\meter}$, the depairing current for this specific sample becomes $I_{\mathrm{dep}} \approx \SI{23}{\milli\ampere}$. We have obtained, thus, that $I_*\sim I_{\mathrm{dep}} \gg I_C$, showing that $I_*$ is governed by the intrinsic depairing mechanism and not by extrinsic defects in this sample.

\subsection{Harmonic Generation}\label{sec:52_HG}
\begin{figure}[t]
    \centering
    % Top panel
    \begin{overpic}[scale=0.4,trim=0 51 0 0,clip]{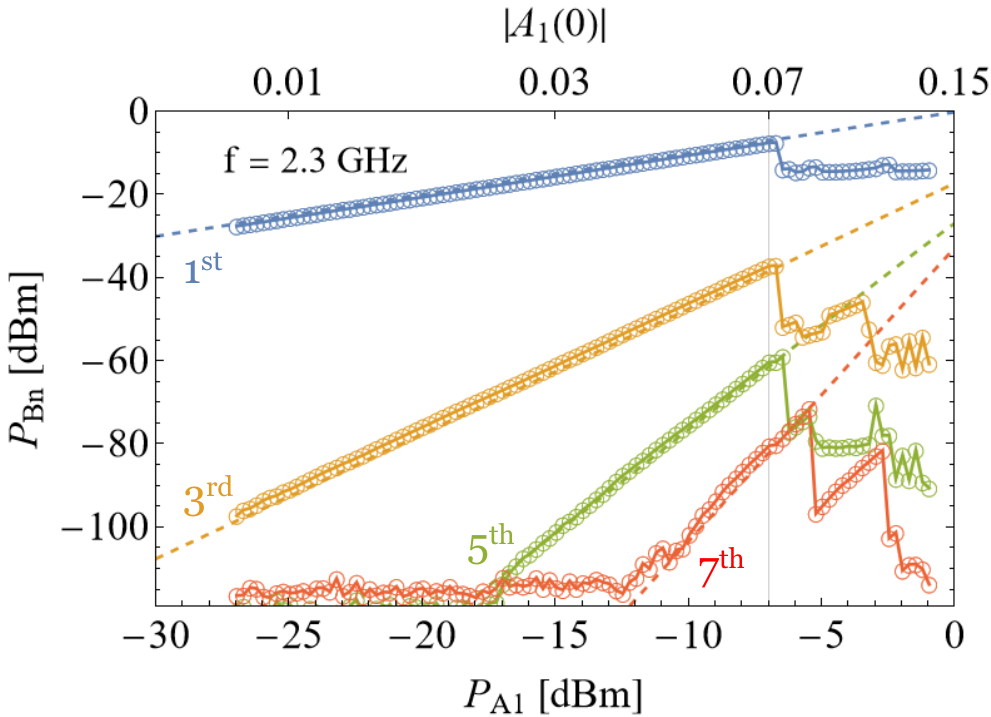}
        \put(0,50){\footnotesize (a)}
    \end{overpic}\\
    % Center panel
    \begin{overpic}[scale=0.4,trim=0 52 0 47,clip]{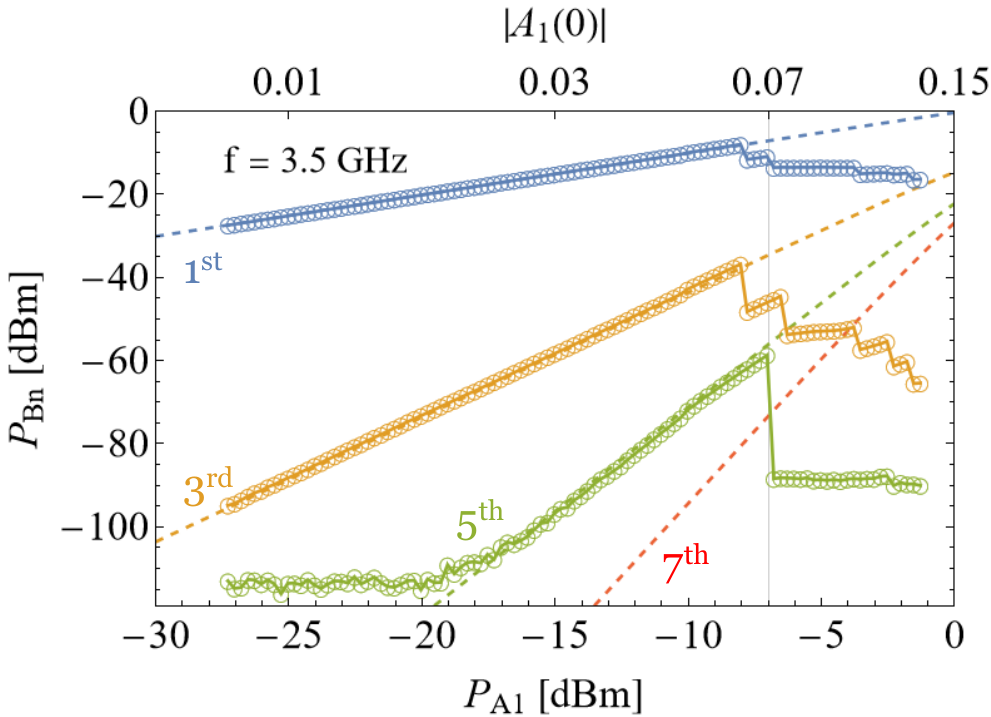}
        \put(0,50){\footnotesize (b)}
    \end{overpic}\\
    % Bottom panel
    \begin{overpic}[scale=0.4,trim=0 0 0 47,clip]{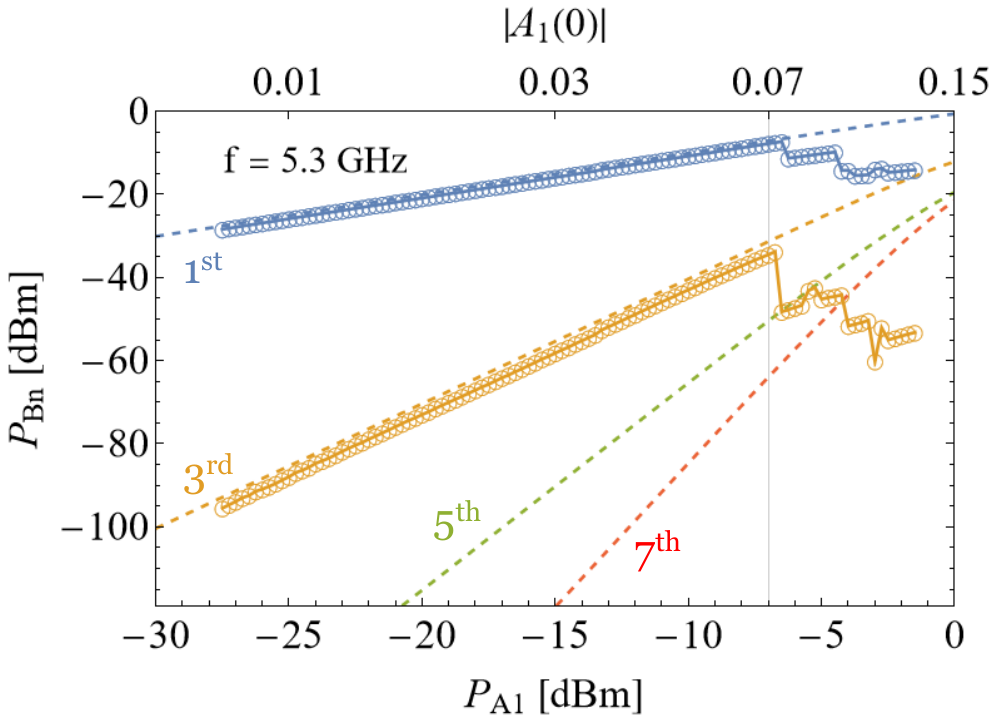}
        \put(0,60){\footnotesize (c)}
    \end{overpic}
    \caption{Harmonic generation at three different fundamental tones. (a)~\SI{2.3}{\giga\hertz}. (b)~\SI{3.5}{\giga\hertz}. (c)~\SI{5.3}{\giga\hertz}. Empty circles are measured data. Where not shown, the harmonics fell outside of the measuring window. Dashed lines are plots of the output powers predicted by the CMEs using $I'_* = \SI{27}{\milli\ampere}$, as obtained from the fit presented in Fig.~\ref{fig:S21_exp}. No fitting was attempted in this case. Within our resolution limit, $\SI{-120}{\dBm}$, we did not see the presence of even harmonics. The gray vertical line represents $P_A = \SI{-7}{\dBm}$ that marks the onset of the first resistive step identified in Fig.~\ref{fig:RI}. The breaking points do not necessarily coincide since every curve was obtained in different sweeps.}
    \label{fig:HG_exp}
\end{figure}
Fig.~\ref{fig:HG_exp} gives three examples of the harmonic generation at different frequencies of the fundamental tone. Every curve was obtained in separate sweeps. Following Fig.~\ref{fig:Model}, we present the output powers at B, $P_{Bn}$, as a function of the incident powers of the fundamental tone at A, $P_A$. For the model presented in \S~\ref{sec:44_HG}, we have used $I'_* = \SI{27}{\milli\ampere}$ as obtained independently from the $S_{21}$ vs. $I_{DC}$ measurements. Notice that we did not attempt any fitting of the model to the experimental data. Up to an incident power of $\sim\SI{-7}{\deci\bel}$, %corresponding to the onset of the first resistive step shown in Fig.~\ref{fig:RI},
a remarkable agreement between experimental data and model is obtained, confirming the validity of the CMEs and the methods used for describing the DE transmission line.

The results presented in Fig.~\ref{fig:HG_exp} also show the presence of dissipative regions where superconductivity is disrupted. They manifest as a sudden decrease in the transmitted power. The entrance to the first step is clearly seen at an incident power of around \SI{-7}{\deci\bel}. As part of the model presented in \S~\ref{sec:44_HG}, this value  can be converted to a normalized input amplitude, giving $A_1(0)=0.07$ which coincides very well with the value of $I_{C1}/I'_*$, further validating our methods.

\subsection{Implications for Design}\label{sec:53_Design}
The fact that $I'_*>I_*\sim I_\mathrm{dep} \gg I_C$ has an important practical consequence, it limits the normalized amplitude with which a traveling-wave parametric device can be fed. This situation is further worsened by the appearance of the dissipative regions whose onset, fortunately, approaches $I_C$ at temperatures much lower than $T_C$. In our sample, at \SI{4}{\kelvin},  the input amplitude is limited to only 7\% of $I'_*$, preventing it from being used as a parametric amplifier when combined with its short length. Besides increasing the length of the transmission line, one practical solution is to increase its kinetic inductance fraction by using thinner superconducting films and decreasing its internal dimensions. In this way, one could achieve $\alpha_K \sim 1$ which, for our sample, maintaining everything else equal and taking into account that $I'_*=I_*/\sqrt{\alpha_K}$, would translate into being able to achieve $A_1(0) \sim 0.09$, nearly a 30\% increase. Moreover, decreasing the film thickness, while maintaining the same fabrication quality, may have the advantage of increasing $I_C$ with respect to $I_\mathrm{dep}$~\cite{Stejic1994}, allowing further increase of $A_1(0)$. Other groups have demonstrated gain~\cite{Eom2012,Faramarzi2024,Malnou2021} with designs consistent with these considerations.

Another important aspect to consider comes from the fact that kinetic-inductance traveling wave devices usually use capacitive stubs to compensate for the high inductance of the line. This geometry is prone to current crowding~\cite{Hortensius2012} further limiting the achievable $I_C$. Rounding properly the intersections between the stubs and the central line may help to increase $I_C$. For our sample at least, we do not know if the critical current is limited by the presence of the stubs or by the central conductor of the CPW line.

%%%%%%%%%%%%%%%%%%%%%%%%%%%%%%%%%%%%%%%%%%%%%%%%%%%%%%%%%%%%%%%%%%%%%%

\section{Conclusions}
We have presented a general and compact expression for obtaining the coupled mode equations for any parametric process in the presence of a quadratic nonlinearity in the inductance of a superconducting transmission line. Furthermore, the expression includes losses which can come not only from material properties but also from the presence of stopbands. We then presented several examples of the use of the equation to illustrate its versatility in representing any combination of traveling waves allowed by the mixing process.

The presented formulation was validated experimentally by studying harmonic generation in a dispersion-engineered superconducting transmission line. Importantly, we performed a parameter-free validation by combining simulations and measurements of the variation of the RF transmission of the line with applied DC current. On the one hand, the simulations allowed us to obtain the parameters of the studied transmission line, its propagation constant and characteristic impedance. On the other hand, the transmission vs. $I_{DC}$ measurements permitted us to determine independently the nonlinear parameter $I'_*$ which normalizes the coupled mode equations. When using the obtained value for $I'_*$ and the simulated properties of the line, a remarkable agreement between theory and experiment was found. Furthermore, we found that $I'_*$ does not scale with the critical current of the sample but with the more fundamental depairing current. This finding has profound implications for the design of traveling-wave parametric devices.

Our findings rigorously confirm the design choices made by other groups when fabricating kinetic-inductance traveling-wave devices, namely using thin superconducting films and long transmission lines. We also underscore the importance of increasing the achievable critical current so as to increase the normalized pump amplitude driving the parametric process inside the device.

%%%%%%%%%%%%%%%%%%%%%%%%%%%%%%%%%%%%%%%%%%%%%%%%%%%%%%%%%%%%%%%%%%%%%%

\section*{Acknowledgments}
We thank Jochem Baselmans (Delft University of Technology and SRON, The Netherlands) for his invaluable support in the fabrication and design of the samples.

R.~Finger and R.~O.~Berriel gratefully acknowledge support of ANID funds Basal FB210003 and ALMA 31240042.

The National Radio Astronomy Observatory and Green Bank Observatory are facilities of the U.S. National Science Foundation operated under cooperative agreement by Associated Universities, Inc.

\section*{Author Declarations}

\subsection*{Conflict of Interest}
The authors have no conflicts to disclose.

\subsection*{Author Contributions}
\textbf{F.~P.~Mena:} Conceptualization (lead); Methodology (lead); 
Software (lead); Formal analysis (lead); Investigation (supporting); 
Writing -- original draft (lead); Writing -- review \& editing (lead).
\textbf{R.~O.~Berriel:} Investigation (lead); Formal analysis (supporting); 
Writing -- review \& editing (supporting).
\textbf{C.~Espinoza:} Investigation (supporting); 
Writing -- review \& editing (supporting).
\textbf{R.~Finger:} Conceptualization (supporting); Methodology (supporting); 
Formal analysis (supporting); Writing -- review \& editing (supporting).
\textbf{D.~J.~Thoen:} Resources (lead); Writing -- review \& editing (supporting).

\section*{Data Availability}
The data that support the findings of this study are available from the corresponding author upon reasonable request.

% \nocite{aipnum4-1Control}
% \bibliographystyle{aipnum4-1}
% \bibliography{references.bib}

\nocite{*}
\bibliography{references.bib}

\end{document}